# A New Hybrid Control Architecture to Attenuate Large Horizontal Wind Disturbance for a Small-Scale Unmanned Helicopter

Regular Paper

Xiaorui Zhu[1,*], Wenwu Zeng[2], Zexiang Li[3] and Chunyang Zheng[4]

1,2,4 School of Mechanical Engineering and Automation, Harbin Institute of Technology Shenzhen Graduate School, China
3 Department of Electronic and Computer Engineering, Hong Kong University of Science and Technology, Hong Kong
* Corresponding author E-mail: hit.zhu.xr@gmail.com

Abstract This paper presents a novel method to attenuate large horizontal wind disturbance for a small-scale unmanned autonomous helicopter combining wind tunnel-based experimental data and a backstepping algorithm. Large horizontal wind disturbance is harmful to autonomous helicopters, especially to small ones because of their low inertia and the high cross-coupling effects among the multiple inputs. In order to achieve more accurate and faster attenuation of large wind disturbance, a new hybrid control architecture is proposed to take advantage of the direct force/moment compensation based on the wind tunnel experimental data. In this architecture, large horizontal wind disturbance is treated as an additional input to the control system instead of a small perturbation around the equilibrium state. A backstepping algorithm is then designed to guarantee the stable convergence of the helicopter to the desired position. The proposed technique is finally evaluated in simulation on the platform, HIROBO Eagle, compared with a traditional wind velocity compensation method.

## 1. Introduction

An unmanned autonomous helicopter (UAH) is suitable for a variety of applications such as surveillance and reconnaissance, search and rescue, urgent cargo transportation, and scientific investigations in some extreme environments. In the last two decades, small-scale UAHs have attracted much more attention because of their small size and agility. However, one of main drawbacks is that a small-scale helicopter is prone to be distracted by large wind disturbance, especially when it is applied outdoors. In order to deal with this issue, this paper proposes a novel method to attenuate the large horizontal wind disturbance combining wind tunnel-based experimental data and a backstepping algorithm. The helicopter platform applied in this paper is the HIROBO Eagle.

In 1991 Carnegie Mellon University first began research on the flight control of small unmanned helicopters [1]. Since then, substantial work has been done on different types of controllers for small UAHs. A few research groups have tried non-model-based control strategies. Some researchers designed fuzzy controllers[2, 3], neural network-based controllers [4] for the unmanned helicopter based on the priori knowledge of the pilot. More recently, a reinforcement learning method was developed for hovering and even aeronautical flying of small-scale unmanned helicopters [5-7]. However, these methods all relied on pilot experience, which is not suitable for many scenarios such as environments with variable weather conditions.

As various modelling approaches were developed, many kinds of model-based control techniques have been developed accordingly to solve the controlling problem for UAH. In the early period, some researchers designed the controllers based on the identified linear models such as the LQG controller with set-point tracking and proportional derivative (PD) controller [8]. Over the last decade, much attention has been paid to nonlinear controllers. Output tracking controllers based on input-output linearization were developed to achieve bounded tracking [9]. A fuzzy gain-scheduling algorithm based on linearization of the nonlinear model [10] was proposed. Some literature also proposed nonlinear predictive controllers [11] [12], neural network controllers based on the nonlinear model [13, 14], $H_\infty$ controllers [15, 16] and nonlinear robust controllers [17, 18] for UAHs. A feedback linearization technique with a Convex Integrated Design (CID) method was proposed [19] for a model helicopter. There has also been much interest in using the backstepping methodology for the design of controllers for UAHs [20-24]. This paper also uses the backstepping method in designing the controller for the small-scale UAH.

However, large horizontal wind disturbance is an important factor affecting the performance of the small-scale helicopters in real outdoor environments where the wind disturbance could not be considered as a small perturbation to the system. So far, only a few studies have focused on the attenuation of the wind disturbance for small-scale aircraft. A minimal order robust controller was proposed to attenuate lateral gusts on an aircraft [25]. An $H_\infty$ controller was designed to dramatically reduce the effect of gusts on aircraft vertical acceleration [26], but they only focused on the airplanes rather than helicopters. More recently, a nonlinear robust controller was developed for a model-scale helicopter to reduce the vertical wind disturbance based on the 3DOF helicopter model [27]. A PD controller was designed for attenuating horizontal gusts [28] where only heave motion dynamics of the helicopter was applied. However, neither of them considered the full dynamics of the helicopter. Considering the 6DOF nonlinear helicopter model, a robust backstepping controller was proposed for helicopters to reject the wind disturbance, taking advantage of the input observer technique to reconstruct wind disturbance [29]. A robust longitudinal lateral and vertical stabilizer was developed for helicopters considering the wind disturbance as a sum of a fixed number of sinusoids with unknown amplitudes, frequencies and phases [30]. Two robust controllers using the approximate feedback linearization procedure were designed to keep stability of the 7DOF model-scale helicopter under lateral and vertical wind gusts [31]. Compared with the above three methods, the novelty of this paper is to introduce the external force/moment estimator from the wind tunnel experiments to the control scheme, which is more realistic. Hence, the main contribution of this paper includes a new hybrid control architecture based on the wind tunnel experimental data and the full dynamics of the helicopter to deal with large horizontal wind disturbances. Moreover, a nonlinear robust controller based on the backstepping method is designed and simulated for a model-scale helicopter to validate the proposed control strategy.

The structure of this paper is organized as follows. Section 2 introduces the nonlinear model of UAH, including the rigid body dynamics, flapping dynamics and servo actuator dynamics. In Section 3, the backstepping controller for small-scale helicopter is developed considering the compensation of wind disturbance using wind tunnel experiment data. In section 4, the simulation and discussion for the proposed method is presented. Finally, the conclusion is made in Section 5.

## 2. Nonlinear model of UAH

The nonlinear model of UAH contains rigid body dynamics (1-3) [32, 33], main rotor and fly-bar flapping dynamics (6-12), and servo actuator dynamics (10-13) [23], Fig. 1. The rigid body dynamics are described as,

$$\dot{\xi} = R^T V \tag{1}$$

$$\dot{V} = -\Omega \times V + \frac{F}{m} + \hat{g} \tag{2}$$

$$J\dot{\Omega} = -\Omega \times J\Omega + M \tag{3}$$

where $\xi = (x, y, z)^T$ is the position vector of the helicopter in the inertial frame; $V = (u, v, w)^T$ is the linear velocity vector of the helicopter in the body frame; $\Omega = (p,q,r)^T$ is angular velocity vector of the helicopter in the body frame; $\hat{g}$ is the gravitational vector in the body frame; $J$, $m$ are the inertial matrix and the mass of the helicopter; $F = [F_x, F_y, F_z]$ and $M = [L_m, R_m, N_m]$ are the vectors of the external forces and moments, respectively.

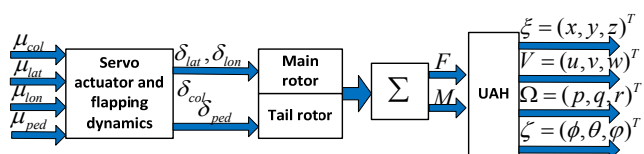

**Figure 1.** The system structure of the helicopter

The gravity vector can be expressed as,

$$\hat{g} = R^T[0,0,g] = [-g\sin\theta \quad g\cos\theta\sin\phi \quad g\cos\theta\cos\phi]^T \quad (4)$$

$$R = \begin{bmatrix} \cos\theta\cos\varphi & \sin\theta\cos\varphi\sin\phi - \sin\varphi\cos\phi & \sin\theta\cos\varphi\cos\phi + \sin\varphi\sin\phi \\ \cos\theta\sin\varphi & \sin\theta\sin\varphi\sin\phi - \cos\varphi\cos\phi & \sin\theta\sin\varphi\cos\phi - \cos\varphi\sin\phi \\ -\sin\theta & \cos\theta\sin\phi & \cos\theta\cos\phi \end{bmatrix} \quad (5)$$

where $(\phi, \theta, \varphi)$ denote the attitude angle and $g$ is the gravity.

The total forces and moments due to the main and tail rotors are given as [32],

$$F = \begin{bmatrix} F_x \\ F_y \\ F_z \end{bmatrix} = \begin{bmatrix} -T_{mr}a_1 \\ T_{mr}b_1 + T_t \\ -T_{mr} \end{bmatrix} \quad (6)$$

$$M = \begin{bmatrix} L_m \\ R_m \\ N_m \end{bmatrix} = \begin{bmatrix} \frac{dL}{db_1}b_1 + F_y k_x + T_t l_z \\ \frac{dM}{da_1}a_1 + F_x k_z \\ M_Q + F_y k_x + T_t l_x \end{bmatrix} \quad (7)$$

where $T_{mr}$, $T_t$ are the main and tail rotor thrusts; $a_1$, $b_1$ are the main rotor flapping angles; the $dL/db_1$, $dL/da_1$ are the rolling and pitching moment stiffness, respectively; $k_x$, $k_z$ are the $x$, $z$ axes' lengths of the main rotor with respect to the centre of the UAH; $l_x$, $l_z$ are the $x$, $z$ axes' lengths of the tail rotor with respect to the centre of the UAH; $M_Q$ is the main rotor torque and the approximate calculation is given as follows [34]:

$$M_Q = C_m T_{mr}^{1.5} + D_m^Q \quad (8)$$

In Eq.(8), $C_m$ and $D_m^Q$ are the torque constants of the main rotor.

The main rotor and fly-bar flapping dynamics are represented as,

$$\dot{a}_1 = -\frac{a_1}{\tau_f} + q + \frac{A_c}{\tau_f}c + \frac{A_{lon}}{\tau_f}\delta_{lon} \quad (9)$$

$$\dot{b}_1 = -\frac{b_1}{\tau_f} + p + \frac{B_d}{\tau_f}c + \frac{B_{lat}}{\tau_f}\delta_{lat} \quad (10)$$

$$\dot{c} = -\frac{c}{\tau_s} + q + \frac{C_{lon}}{\tau_s}\delta_{lon} \quad (11)$$

$$\dot{d} = -\frac{d}{\tau_s} + p + \frac{D_{lat}}{\tau_s}\delta_{lat} \quad (12)$$

where $\delta_{lon}$, $\delta_{lat}$, $\delta_{ped}$, $\delta_{col}$ are the servo actuator outputs and $c$, $d$ are the fly-bar flapping angles. The $\tau_f$, $\tau_s$ are the flapping time constants of the main rotor and tail rotor, respectively. The parameters $A_c$, $A_{lon}$, $B_d$, $B_{lat}$, $C_{lon}$, $D_{lat}$ are the constant coefficients for a specific type of the helicopter.

The first-order servo actuator model is given by,

$$\dot{\delta}_{lon} = -\tau_{lon}\delta_{lon} + u_{lon} \quad (13)$$

$$\dot{\delta}_{lat} = -\tau_{lat}\delta_{lat} + u_{lat} \quad (14)$$

$$\delta_{col} = K_{col}u_{col} \quad (15)$$

$$\delta_{ped} = K_{ped}u_{ped} \quad (16)$$

where $u_{lon}$, $u_{lat}$, $u_{col}$, $u_{ped}$ are the servo actuator inputs and $\tau_{lon}$, $\tau_{lat}$, $K_{col}$, $K_{ped}$ are the constant coefficients for a specific type of the helicopters.

There is an approximate algebraic relationship between $T_{mr}$ and $\delta_{col}$ [35],

$$\begin{cases} T_{mr} = (w + A_0\delta_{col} - V_{imr})\frac{B_0}{4} \\ V_{imr}^2 = \sqrt{\frac{1}{4}\hat{V}_{mr}^2 + C_0 T_{mr}^2} - \frac{1}{2}\hat{V}_{mr}^2 \\ \hat{V}_{mr} = u^2 + v^2 + w(w - 2V_{imr}) \end{cases} \quad (17)$$

The relative parameters are given as,

$$A_0 = \frac{2}{3}\Omega_{mr}R_{mr}, \quad B_0 = \rho\Omega_{mr}R_{mr}^2 A_{mr} B_{mr} c_{mr}, \quad C_0 = \frac{1}{4\rho^2\pi^2 R_{mr}^4}$$

where $R_{mr}$ is the radius of the main rotor and $\Omega_{mr}$ is the angular velocity of the main rotor. The parameter $B_{mr}$ denotes the number of blades and $c_{mr}$ is the blade chord of the main rotor, $\rho$ is the density of the air. Similarly, an approximate algebraic relationship lies in $T_t$ and $\delta_{ped}$ [35],

$$\begin{cases} T_t = (v + rT_x + pT_z + A_1\delta_{ped} - V_{it})\frac{B_1}{4} \\ V_{it}^2 = \sqrt{\frac{1}{4}\hat{V}_t^2 + C_1 T_t^2} - \frac{1}{2}\hat{V}_t^2 \\ \hat{V}_t = u^2 + v^2 + w(w - 2V_{it}) \end{cases} \quad (18)$$

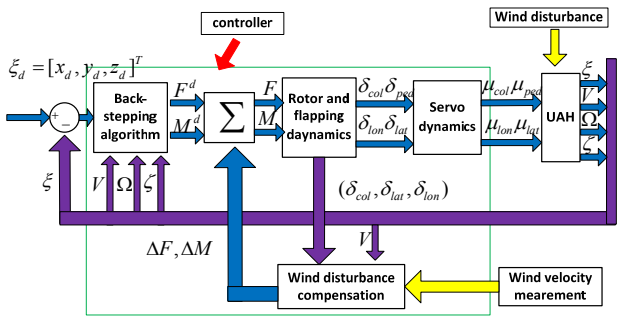

**Figure 2.** Block diagram of the proposed control architecture. The relative parameters are given as,

$$A_1 = \frac{2}{3}\Omega_t R_t, \quad B_1 = \rho \Omega_t R_t^2 A_t B_t c_t, \quad C_1 = \frac{1}{4\rho^2 \pi^2 R_t^4}$$

where $R_t$ represents the radius of the tail rotor and $\Omega_t$ is the angular velocity of the tail rotor. The parameter $B_t$ denotes the number of blades and $c_t$ is the blade chord of the tail rotor.

## 3. Control System Design

### 3.1 Hybrid Control Architecture

The hybrid control architecture is proposed to achieve more accurate and faster attenuation of large horizontal wind disturbance for a small-scale helicopter, Fig. 2. In this paper, a three dimensional position vector of the helicopter, $\xi_d = (x_d, y_d, z_d)^T$, is given as a reference input. The desired force/moment vector ($F^d$, $M^d$) is designed using a backstepping algorithm without considering the wind disturbance. The wind velocity $\delta_{wind}$ is treated as an additional input to the control system instead of being treated as a small perturbation in the case of small horizontal wind disturbances. According to the wind tunnel experiment data, a force/moment vector ($\Delta F$, $\Delta M$) is designed to compensate for the wind disturbance. This compensation is then fed into a backstepping algorithm to guarantee that the small-scale UAH can stably approach the desired position in a finite time.

### 3.2 Force/Moment Estimation for Wind Disturbance

In order to compensate for the force/moment caused by large horizontal wind disturbance, a quantities relationship between the wind velocity and the force/moment needs to be estimated based on the experimental data obtained from the wind tunnel experiments. It is difficult to obtain an explicit solution theoretically, however, a polynomial formulation could be used to estimate the relationship between the force/moment and the control inputs (collective/longitude/lateral/wind velocity). In the following paragraphs, three polynomial functions are derived to estimate the forces in different axes caused by large horizontal wind disturbance. Note that the resulting moments can be obtained easily from the force estimations. Hence, the moment expression is skipped in this section.

According to the first principles of the helicopter, the thrust force depends on the collective input $\delta_{col}$ of the main rotor. As observed during the wind tunnel experiments, the thrust force is closely related to the wind velocity. In the condition of hovering, the thrust completely projects to the $z$ axis. Therefore, the force $F_z$ applied on the helicopter can be estimated as the function of the collective input $\delta_{col}$, and the wind velocity $\delta_{wind}$:

$$F_z = p_6 \cdot \delta_{col}^6 \delta_{wind}^6 + p_5 \cdot \delta_{col}^5 \delta_{wind}^5 + \cdots + p_0 \cdot \delta_{col}^0 \delta_{wind}^0 \quad (19)$$

where $p_i$ ($i = 0 \cdots 6$) represent the coefficients of the polynomial function.

The applied forces in $x$ and $y$ axes are cross-related with the longitude input and the lateral input of the main rotor due to the aerodynamics. According to the wind tunnel experiments, these forces in $x$ and $y$ axes are also affected by the wind velocity. Hence, forces $F_x$ and $F_y$ can be estimated as the function of the longitude cyclic pitch input $\delta_{lon}$, lateral cyclic pitch input $\delta_{lat}$ and the wind velocity $\delta_{wind}$:

$$\begin{aligned} F_x &= a_{666}\delta_{lat}^6\delta_{lon}^6\delta_{wind}^6 + a_{665}\delta_{lat}^6\delta_{lon}^6\delta_{wind}^5 + \cdots + a_{660}\delta_{lat}^6\delta_{lon}^6\delta_{wind}^0 \\ &+ a_{656}\delta_{lat}^6\delta_{lon}^5\delta_{wind}^6 + a_{655}\delta_{lat}^6\delta_{lon}^5\delta_{wind}^5 + \cdots + a_{650}\delta_{lat}^6\delta_{lon}^5\delta_{wind}^0 + \cdots \\ &+ a_{006}\delta_{lat}^0 \cdot \delta_{lon}^0\delta_{wind}^6 + a_{005}\delta_{lat}^0\delta_{lon}^0\delta_{wind}^5 + \cdots + a_{000}\delta_{lat}^0\delta_{lon}^0\delta_{wind}^0 \end{aligned} \quad (20)$$

$$\begin{aligned} F_y &= b_{666}\delta_{lat}^6\delta_{lon}^6\delta_{wind}^6 + b_{665}\delta_{lat}^6\delta_{lon}^6\delta_{wind}^5 + \cdots + b_{660}\delta_{lat}^6\delta_{lon}^6\delta_{wind}^0 \\ &+ b_{656}\delta_{lat}^6\delta_{lon}^5\delta_{wind}^6 + b_{655}\delta_{lat}^6\delta_{lon}^5\delta_{wind}^5 + \cdots + b_{650}\delta_{lat}^6\delta_{lon}^5\delta_{wind}^0 + \cdots \\ &+ b_{006}\delta_{lat}^0\delta_{lon}^0\delta_{wind}^6 + b_{005}\delta_{lat}^0\delta_{lon}^0\delta_{wind}^5 + \cdots + b_{000}\delta_{lat}^0\delta_{lon}^0\delta_{wind}^0 \end{aligned} \quad (21)$$

where $a_{iii}(i=0\cdots 6)$ denotes the coefficient of the polynomial function $F_x$ and $b_{iii}(i=0\cdots 6)$ is the coefficient of the polynomial function $F_y$.

The wind tunnel experiment data can provide a number of the inputs ($\delta_{lon}$, $\delta_{lat}$, $\delta_{col}$), the wind velocity $\delta_{wind}$ and the resulting forces $F_x$, $F_y$, $F_z$. Based on these data, all the coefficients in Eq. (19)-(21) can be identified using the least square method.

### 3.3 Backstepping Control with Force/Moment Compensation

In the proposed architecture, a backstepping algorithm is applied to incorporate the Lyapunov redesign and the force/moment compensation such that the desired forces and moments could be generated for the helicopter with large horizontal wind disturbances, Fig. 2. The three di-mensional position $\xi$, the attitude $\zeta$, the linear velocities $V$, the angular velocities $\Omega$ of the helicopter and the wind velocity can be measured and fed back to the controller in this paper. Four control inputs ($u_{col}$, $u_{ped}$, $u_{lat}$, $u_{lon}$) will be designed to achieve posture regulation of the helicopter. The Lyapunov redesign is extended from the method in Ahmed et al. [23].

The first Lyapunov function is chosen as,

$$W_1 = \frac{1}{2}(\xi - \xi_d)^T(\xi - \xi_d) \qquad (22)$$

where $\xi_d$ is the desired position. Then the time derivative of Eq. (22) is obtained as,

$$\dot{W}_1 = (\xi - \xi_d)^T V \qquad (23)$$

If the desired velocity, $V^d$, satisfies the following equation:

$$V^d = -\alpha(\xi - \xi_d) \qquad (24)$$

where $\alpha$ ($\alpha \geq 0$) is a controller parameter and the actual velocity is equal to the desired velocity, then we can get

$$\dot{W}_1 = -\alpha(\xi - \xi_d)^T(\xi - \xi_d) \leq 0 \qquad (25)$$

If the error between the actual velocity and the desired velocity is denoted as $z_1 = V - V^d$, the second Lyapunov function is defined as,

$$W_2 = \frac{1}{2}(\xi - \xi_d)^T(\xi - \xi_d) + \frac{1}{2}z_1^T z_1 \qquad (26)$$

Its time derivative is obtained as,

$$\dot{W}_2 = (\xi - \xi_d)^T V^d + (\xi - \xi_d)^T z_1 + \alpha z_1^T z_1 + \alpha z_1^T V^d \\ + z_1^T (V \times \Omega + \frac{F^d}{m} + \hat{g}) \qquad (27)$$

Then another controller parameter $\beta$ is introduced to satisfy the following equation:

$$V \times \Omega = -\alpha V - (\xi - \xi_d) - \beta \Omega \qquad (28)$$

In order to guarantee $\dot{W}_2 < 0$, $F^d$ is chosen such that:

$$\begin{cases} z_1^T \frac{F^d}{m} = z_1^T(\beta\Omega - \hat{g})(z_1 \neq 0) \\ \frac{F^d}{m} = -\hat{g}(z_1 = 0) \end{cases} \qquad (29)$$

Substituting Eq. (28) and (29) into Eq.(27), the derivative of the second Lyapunov function becomes:

$$\dot{W}_2 = -\alpha(\xi - \xi_d)^T(\xi - \xi_d) \leq 0 \qquad (30)$$

Then the desired angular velocity $\Omega^d$ is defined as,

$$\Omega^d = (S(V) + \beta I)^{-1}(-\alpha V - (\xi - \xi_d)) \qquad (31)$$

where

$$S(V) = \hat{V} = \begin{bmatrix} 0 & -w & v \\ w & 0 & -u \\ v & u & 0 \end{bmatrix} \qquad (32)$$

Define the error between the actual and the desired angular velocity as $z_2 = \Omega - \Omega^d$ and choose the third Lyapunov function as,

$$W_3 = \frac{1}{2}(\xi - \xi_d)^T(\xi - \xi_d) + \frac{1}{2}z_1^T z_1 + \frac{1}{2}z_2^T z_2 \qquad (33)$$

Then the time derivative of $W_3$ becomes:

$$\dot{W}_3 = (\xi - \xi_d)^T V^d + z_2^T(\dot{\Omega} - z_2^T \dot{\Omega}^d + S(V^d)^T z_1) \qquad (34)$$

To make $\dot{W}_3 \leq 0$, we have the following condition:

$$\dot{\Omega} - z_2^T \dot{\Omega}^d + S(V^d)^T z_1 = 0 \qquad (35)$$

Then substitute (35) into (3), and the desired $M^d$ can be represented as,

$$M^d = J(\dot{\Omega}^d - S^T(V^d)V) + \Omega \times J\Omega \qquad (36)$$

Up to now, we have obtained the desired force (29) and desired moment (36) via the backstepping algorithm without considering the wind disturbances.

According to Section 3.2, the compensated force and moment caused by the wind disturbance can be estimated by several polynomial functions. Then we add these compensations to the desired force and moment to generate the new reference forces and moments under the wind disturbance. Assume that the corresponding forces without wind disturbance in three axes from Eq.(19)-(21) are denoted as,

$$\begin{cases} Fx = F_x(\delta_{lon}, \delta_{lat}, 0) \\ Fy = F_y(\delta_{lon}, \delta_{lat}, 0) \\ Fz = F_z(\delta_{col}, 0) \end{cases} \qquad (37)$$

where $Fx$, $Fy$, $Fz$ represent the external forces without the wind disturbance. The forces on three axes under the wind disturbance from Eq. (19)-(21) are denoted as,

$$\begin{cases} Fx_w = F_x(\delta_{lon}, \delta_{lat}, \delta_{wind}) \\ Fy_w = F_y(\delta_{lon}, \delta_{lat}, \delta_{wind}) \\ Fz_w = F_z(\delta_{col}, \delta_{wind}) \end{cases} \qquad (38)$$

So the changes of the forces $\Delta F$ caused by the wind disturbance are represented as,

$$\Delta F = \begin{bmatrix} Fx_w - Fx & Fy_w - Fy & Fz_w - Fz \end{bmatrix} \quad (39)$$

The moment around the $z$ axis is usually attenuated by a self-balancing apparatus on the tail of the UAH. Hence, we only focus on the attenuation of the moments around the $x$ and $y$ axes under the wind disturbance. Assume that the corresponding moments $Mx_w$ and $My_w$ on the $x$ and $y$ axes under the wind disturbance are denoted as,

$$\begin{cases} Mx_w = M_x(\delta_{lon}, \delta_{lat}, \delta_{wind}) \\ My_w = M_y(\delta_{lon}, \delta_{lat}, \delta_{wind}) \end{cases} \quad (40)$$

where $M_x$, $M_y$ denote the polynomial functions between the moments and the inputs $\delta_{lon}$, $\delta_{lat}$, $\delta_{col}$ under the experiment wind tunnel. The moments $Mx$ and $My$ on the $x$ and $y$ axes without wind disturbance are represented as,

$$\begin{cases} Mx = M_x(\delta_{lon}, \delta_{lat}, 0) \\ My = M_y(\delta_{lon}, \delta_{lat}, 0) \end{cases} \quad (41)$$

The change of the moments $\Delta M$, therefore, can be calculated as,

$$\Delta M = \begin{bmatrix} Mx_w - Mx & My_w - My & 0 \end{bmatrix} \quad (42)$$

Hence the reference forces and moments can be described as,

$$F = F^d + \Delta F \quad (43)$$

$$M = M^d + \Delta M \quad (44)$$

where $F$ and $M$ is the modified force and moment considering the wind disturbance. Substituting Eq. (43) and (44) into Eq. (6)-(7), numerical solutions of the desired values of $T_{mr}$, $T_t$, $a_1$, $b_1$ can be calculated using the Gauss–Newton method. Then we can get the control inputs substituting the modified values into Eq. (9)-(16) and Eq.(17)-(18).

3.4 Control Parameter Bounds

According to the previous section, there are two control parameters, $\alpha$ and $\beta$, which affect the performance of the control system. Therefore, the bounds of these control parameters are derived in this section.

From Eq. (24)-(25), it is easy to know the first constrain $\alpha \geq 0$. Substituting (4) and (5) into , we can get:

$$\begin{cases} \beta p + g \sin \theta = -\dfrac{T_{mr} a_1}{m} \\ \beta q - g \cos \theta \sin \phi = \dfrac{T_{mr} b_1 + T_t}{m} \\ \beta r - g \cos \theta \cos \phi = -\dfrac{T_{mr}}{m} \end{cases} \quad (45)$$

From (45), we have the equation:

$$|\beta| = \dfrac{\left\| \dfrac{F}{m} + \hat{g} \right\|}{\|\Omega\|} = \dfrac{\sqrt{X^2 + Y^2 + Z^2}}{\sqrt{p^2 + q^2 + r^2}} \quad (46)$$

where,

$$\begin{cases} X = -\dfrac{T_{mr} a_1}{m} - g \sin \theta \\ Y = \dfrac{T_{mr} b_1 + T_t}{m} + g \cos \theta \sin \phi \\ Z = -\dfrac{T_{mr}}{m} + g \cos \theta \cos \phi \end{cases} \quad (47)$$

According to Eq. (47), we have

$$\begin{cases} X^2 \leq 2(\dfrac{T_{mr}^2 a_1^2}{m^2}) + 2g^2 \sin^2 \theta \\ Y^2 \leq 4 \dfrac{T_{mr}^2 b_1^2 + T_t^2}{m^2} + 2g^2 \cos^2 \theta \sin^2 \phi \\ Z^2 \leq \dfrac{T_{mr}^2}{m^2} + g^2 \cos^2 \theta \cos^2 \phi \end{cases} \quad (48)$$

Let $|T_{mr}| \leq T_0$, $|T_t| \leq t_0$, $|a_1| \leq a_0$, $|b_1| \leq b_0$, $|\theta| \leq \theta_0$, $|\phi| \leq \phi_0$, and the values of $\theta$, $\phi$ are small enough, then we can get:

$$\begin{aligned} X^2 &\leq 2(\dfrac{T_0^2 a_0^2}{m^2}) + 2g^2 \theta_0^2 = X_m^2 \\ Y^2 &\leq 4 \dfrac{T_0^2 b_0^2 + t_0^2}{m^2} + 2g^2 \phi_0^2 = Y_m^2 \\ Z^2 &\leq 2\dfrac{T_0^2}{m^2} + 2g^2 \phi_0^2 = Z_m^2 \end{aligned} \quad (49)$$

According to (45)-(49), the bound of the parameter $\beta$ is given as,

$$|\beta| \leq \dfrac{\sqrt{X_m^2 + Y_m^2 + Z_m^2}}{\|\Omega\|_{\min}} \quad (50)$$

Then Eq. (28) is rewritten as,

$$\alpha V = -V \times \Omega - (\xi - \xi_d) - \beta \Omega \quad (51)$$

Let $u_0 \leq \|V\| \leq u_1$, $w_0 \leq \|\Omega\| \leq w_1$, $\|\xi\| \leq P$, the following inequalities are derived as,

$$\alpha = \dfrac{\|-V \times \Omega - (\xi - \xi_d) - \beta \cdot \Omega\|}{\|V\|} \leq \dfrac{\|-V \times \Omega\| + \|\xi\| + |\beta| \cdot \|\Omega\|}{\|V\|} \quad (52)$$

$$\alpha \leq \dfrac{\|V\| \cdot \|\Omega\| + \|\xi\| + |\beta| \cdot \|\Omega\|}{\|V\|} \leq \dfrac{u_1 \cdot w_1 + P + |\beta|_{\max} \cdot w_1}{u_0} \quad (53)$$

From (49),

$$|\beta|_{\max} = \dfrac{\sqrt{X_m^2 + Y_m^2 + Z_m^2}}{w_0} \quad (54)$$

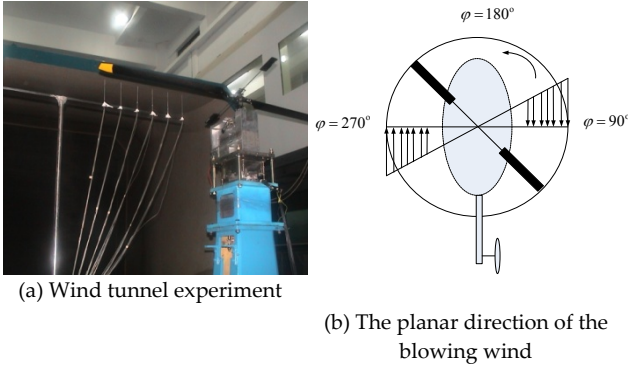

(a) Wind tunnel experiment

(b) The planar direction of the blowing wind

**Figure 3.** Experiment setups.

To sum up, the parameter $\alpha$ satisfies:

$$0 \leq \alpha \leq \frac{u_1 \cdot w_1 + P + |\beta|_{max} \cdot w_1}{u_0} \quad (55)$$

## 4. Simulation and Discussion

### 4.1 Procedures and Parameters

According to the design procedures in Section 3, the proposed control system was simulated in MATALAB/SIMULINK@. All parameters of the platform HILOBO eagle-2 are shown in Table 1 [23]. In order to numerically estimate the relationship between the force/moment on the small-scale helicopter and the wind disturbances, a group of wind tunnel experiments were conducted at the National Lab of Helicopter Dynamics at Nanjing University of Aeronautics and Astronautics, Fig.3 (a). The forces and moments were measured by a 6-DOF balance with 0.1% accuracy and ranged from 0-30(kg) under different wind velocities and directions. The wind velocity varied from 2m/s to 8m/s, the range of our concern. The wind was blown horizontally from different directions of the fuselage such as 0 (longitude) and 270 (lateral) degrees respectively, Fig.3 (b).

Notice that the proposed technique is based on the force/moment compensation for large horizontal wind disturbance. In order to do a comparison, a control system based on the velocity compensation, Fig.4, is introduced where the Dryden wind turbulence is the traditional velocity compensation fed into the undisturbed system. The Dryden wind turbulence model has been widely used to study the response of aircraft to turbulence in the field of flight mechanics. The Dryden wind turbulence model uses the Dryden spectral representation and the forming filters to estimate wind turbulences [36]. The transfer function of the forming filters is shown in Table 2, where $s$ is the white noise with the noise power value $\pi$, $L_w$, $L_u$, $L_v$ represent the turbulence scale lengths, $\sigma_u$, $\sigma_v$, $\sigma_w$ are the turbulence intensities, $h$ is the altitude in feet and $V_{wind}$ is the wind speed at the 20-foot altitude.

| No. | Parameter (Units) | No. | Parameter (Units) |
|---|---|---|---|
| 1 | $m = 7.6(kg)$ | 10 | $J_{xx} = 0.23, J_{yy} = 0.82, J_{zz} = 0.4(kgm^2)$ |
| 2 | $k_x = 0 \; k_z = -0.284(m)$ | 11 | $\Omega_{mr} = 167.55, \Omega_t = 884.3(rad/s)$ |
| 3 | $l_x = -0.915, l_z = -0.104(m)$ | 12 | $R_{mr} = 0.84, R_t = 0.13(m)$ |
| 4 | $\tau_{lon} = 0.04s, \tau_{lat} = 0.04s$ | 13 | $A_{mr} = 5.7, A_t = 4.0(1/rad)$ |
| 5 | $\tau_f = 0.0278, \tau_s = 0.22686(s)$ | 14 | $B_{mr} = 2, B_t = 2$ |
| 6 | $A_c = 0.152, B_d = 0.136(rad/ms)$ | 15 | $C_{mr} = 0.058, C_t = 0.026(m)$ |
| 7 | $A_{lon} = 0.19, B_{lat} = 0.17(rad/ms)$ | 16 | $\rho = 1.225(kg/m^3)$ |
| 8 | $C_{lon} = 1.58, D_{lat} = 1.02(rad/ms)$ | 17 | $dL/db_1 = 199.7(Nm/rad)$ |
| 9 | $C_m = 0.0044, D_m^Q = 0.6304$ | 18 | $dL/da_1 = 107.4(Nm/rad)$ |

**Table 1.** The parameter value of the HILOBOeagle-2 [23].

| Direction | Transfer function | Parameter value |
|---|---|---|
| Longitudinal $D_u(s)$ | $\sigma_u \sqrt{\frac{2L_u}{\pi U}} \frac{1}{1 + \frac{L_u}{U} s}$ | $2L_w = h$ $L_u = 2L_v = \frac{h}{(0.177 + 0.000823h)^{1.2}}$ |
| Lateral $D_v(s)$ | $\sigma_v \sqrt{\frac{2L_v}{\pi U}} \frac{1 + \frac{2\sqrt{3}L_v}{U} s}{(1 + \frac{2L_v}{U} s)^2}$ | $U = \sqrt{u^2 + v^2}$ $\sigma_u = \sigma_v = \frac{\sigma_w}{(0.177 + 0.000823h)^{0.4}}$ $\sigma_w = 0.1 V_{wind}$ |

**Table 2.** Transfer function of the forming filter (MIL-HDBK-1797).

Two different groups of wind disturbances will be simulated to test the performance of the control system. As shown in Fig. 5 (Type I), there is no wind disturbance at the beginning of 10s and the last 10s, and there is a constant wind velocity (varying from 2 m/s to 8 m/s) during 10s and 20s. As shown in Fig. 5 (Type II), wind disturbances exist during the whole simulation period (30s) and the wind velocity is 2 m/s, 4 m/s, 6 m/s, 8 m/s, respectively. In the simulation, the initial position is set at (5m,-5m,-5m) while the desired position is the origin.

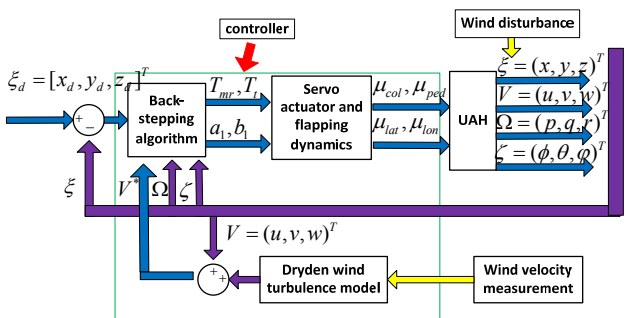

**Figure 4.** The controller block based on the Dryden wind turbulence model.

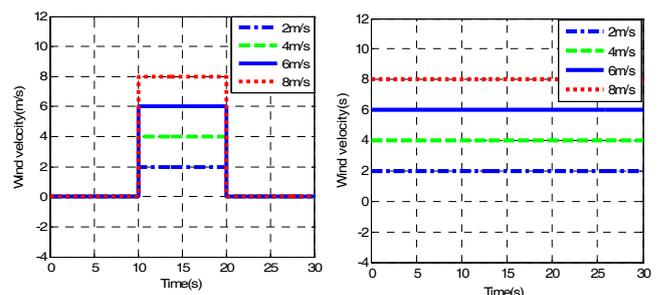

(a) Type I         (b) Type II

**Figure 5.** Wind velocity inputs.

4.2 Simulation Results and Discussion

**Case A:** In Case A, Type I wind disturbances were applied to simulate a short-period gust wind from 0 (longitude) and 270 (lateral) degree directions, respectively. In this section, solid lines, dotted lines and dash-dotted lines represent the $x$-axis, $y$-axis and $z$-axis positions, respectively. Fig.6 shows the position regulation of the helicopter based on the velocity compensation where the parameters are chosen as $\alpha$ =2.5, $\beta$ =3 for better performance in the conditions of different longitude wind speeds (2m/s~8m/s). Fig.7 shows the position regulation result based on the proposed force/moment compensation method where the parameters are chosen as $\alpha$ =4, $\beta$ =2 for better performance in the same range of longitude wind speeds. Fig.8 shows the control inputs of the proposed force/moment compensation method when the longitude velocity of short-period wind is 6m/s. According to the above results, we can find out that two position controllers both converge to the desired position finally. However, the oscillation is larger and the convergence is 10s slower using the traditional velocity compensation based on the Dryden wind turbulence model in dealing with the gust of wind.

Lateral wind disturbance was then applied to verify the system performance. Fig.9 shows the position regulation of the helicopter based on the velocity compensation where the parameters are chosen as $\alpha$ =2, $\beta$ =2.5 for better performance in the conditions of different lateral wind speeds (2m/s~8m/s). Fig.10 shows the position regulation results based on the proposed force/moment compensation method where the parameters are chosen as $\alpha$ =2.5, $\beta$ =3 for better performance under the same range of lateral wind speeds. Fig.11 shows the control inputs of the proposed force/moment compensation method when the lateral velocity of short-period wind is 6m/s. Comparing Fig. 9 and Fig. 10, it is easy to conclude that the proposed

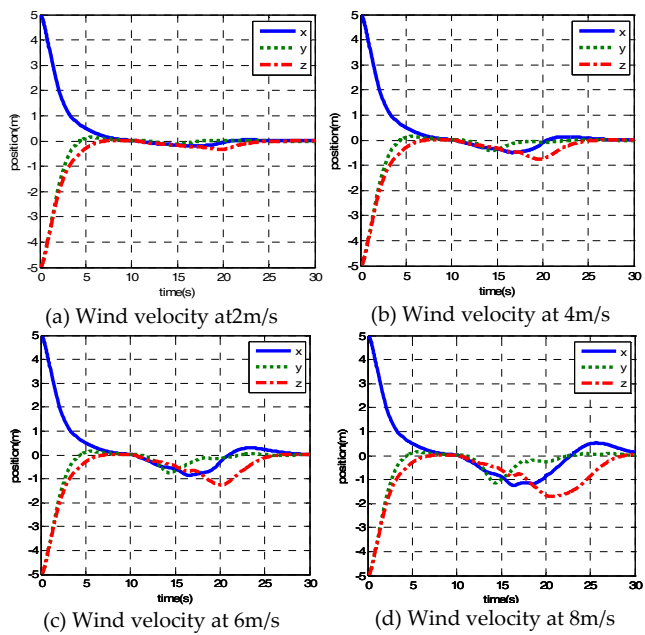

**Figure 6.** Performance of the position control based on velocity compensation in Case A with longitude wind disturbance.

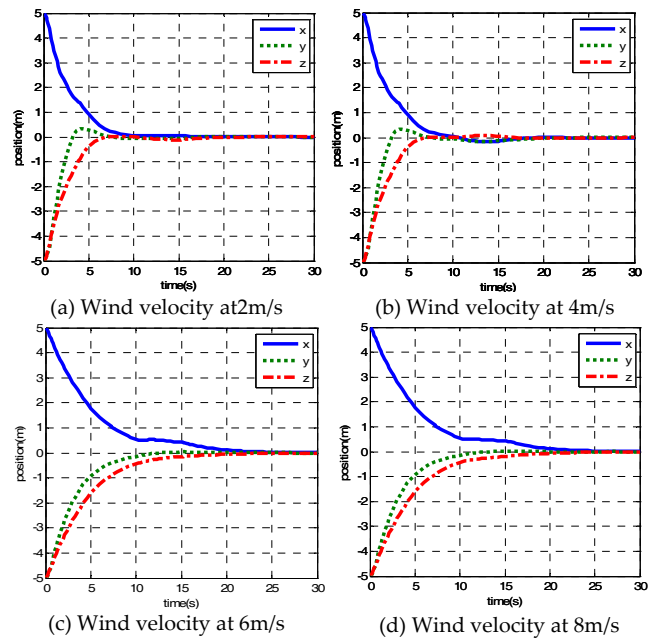

**Figure 7.** Performance of the position control based on the proposed method in Case A with longitude wind disturbance.

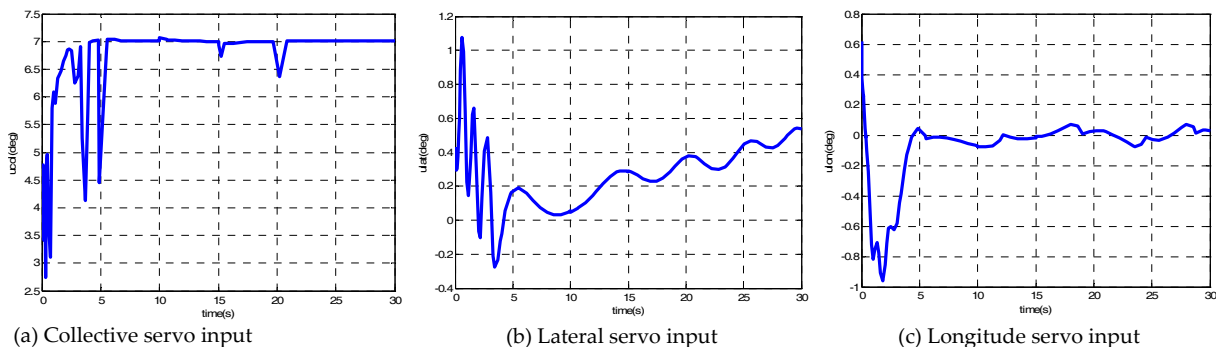

**Figure 8.** Servo actuator inputs for the proposed method at the 6m/s wind velocity of Case A from the longitude direction.

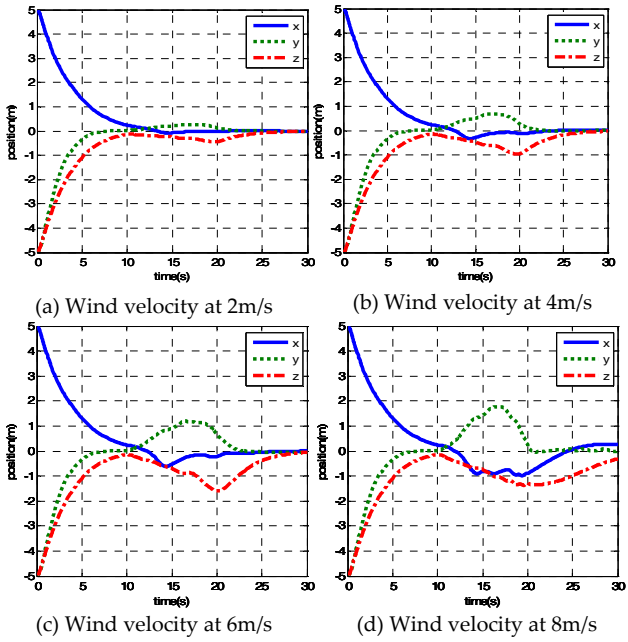

**Figure 9.** Position regulation results based on velocity compensation in Case A with lateral wind disturbance.

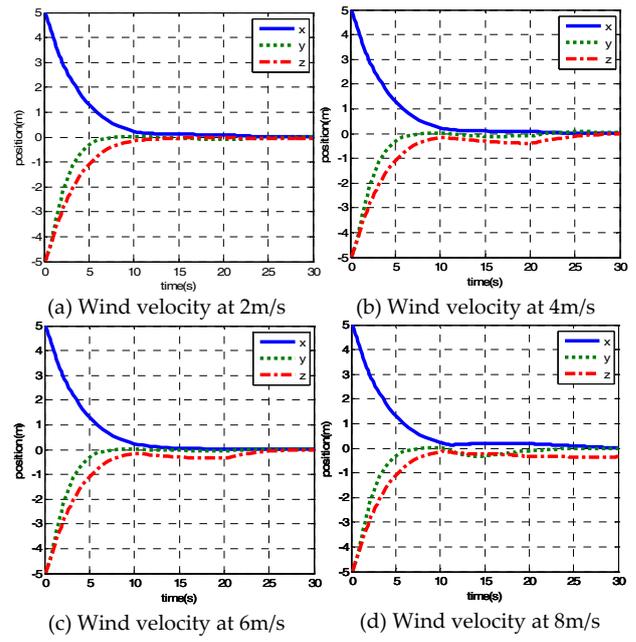

**Figure 10.** Position regulation results based on the proposed method in Case A with lateral wind disturbance.

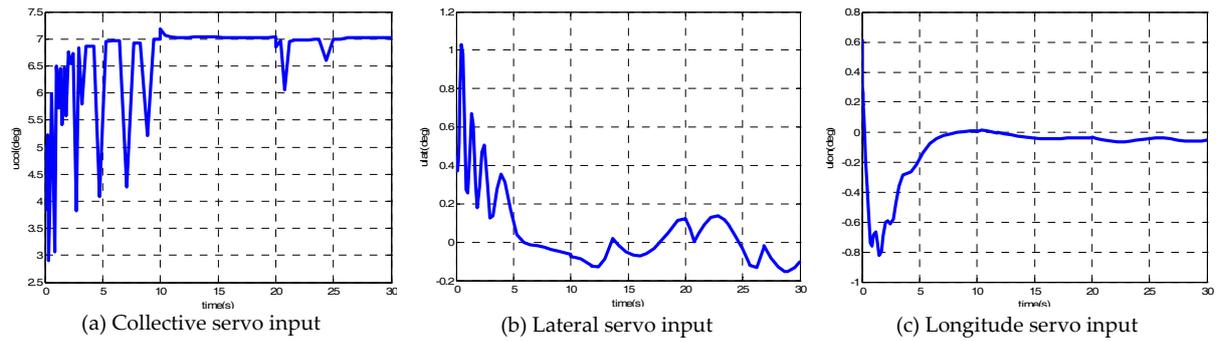

**Figure 11.** Servo actuator inputs for the proposed method at the 6m/s wind velocity of Case A from the lateral direction.

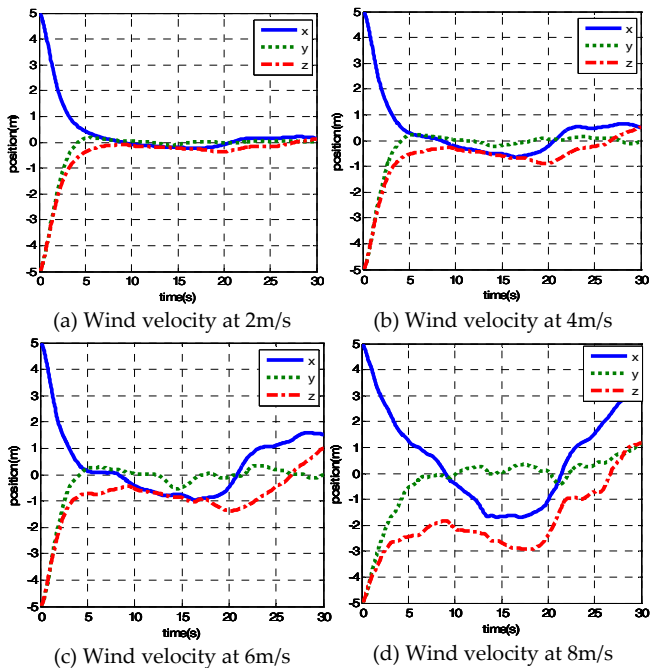

**Figure 12.** Position regulation results based on velocity compensation in Case B with longitude wind disturbance.

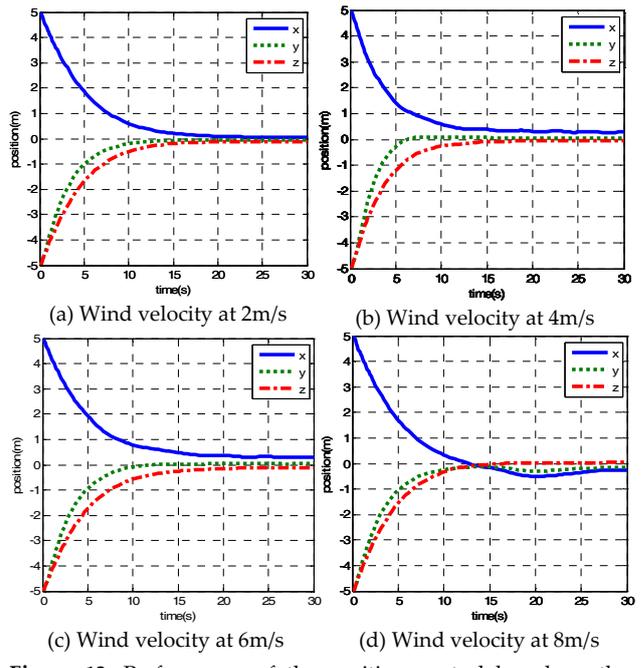

**Figure 13.** Performance of the position control based on the proposed method in case B with longitude wind disturbance.

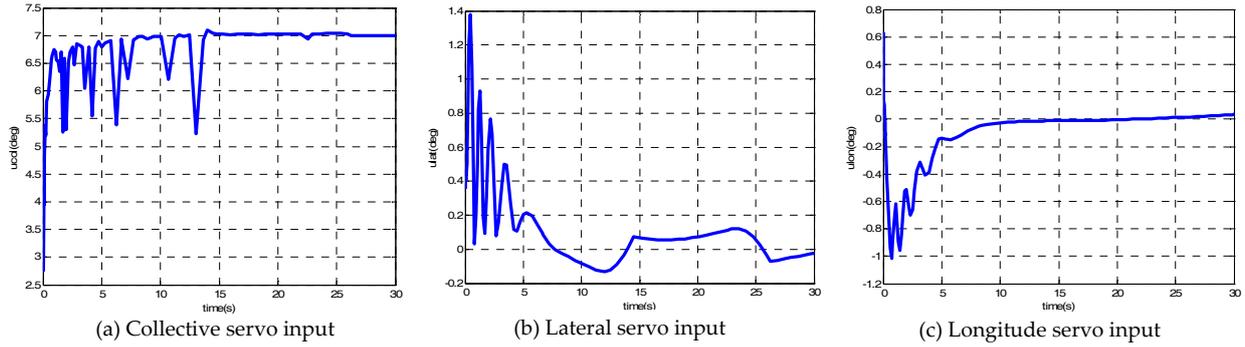

(a) Collective servo input  (b) Lateral servo input  (c) Longitude servo input

**Figure 14.** Servo actuator inputs for the proposed method at the 6m/s wind velocity of Case B from the longitude direction.

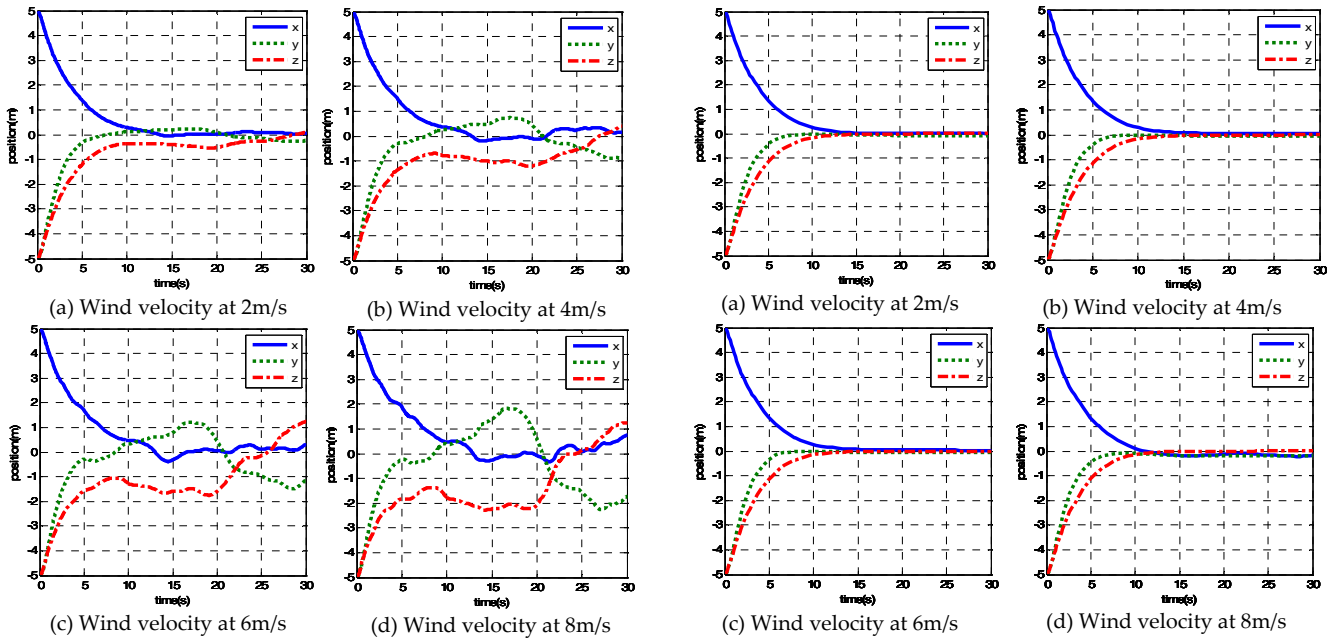

(a) Wind velocity at 2m/s  (b) Wind velocity at 4m/s  (a) Wind velocity at 2m/s  (b) Wind velocity at 4m/s

(c) Wind velocity at 6m/s  (d) Wind velocity at 8m/s  (c) Wind velocity at 6m/s  (d) Wind velocity at 8m/s

**Figure 15.** Position regulation results based on velocity compensation in Case B with lateral wind disturbance.

**Figure 16.** Position regulation results based on the proposed method in Case B with lateral wind disturbance.

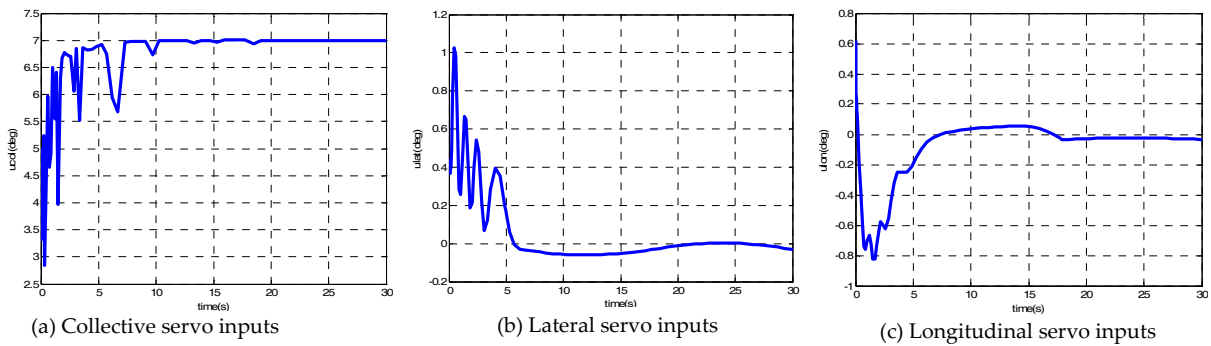

(a) Collective servo inputs  (b) Lateral servo inputs  (c) Longitudinal servo inputs

**Figure 17.** Servo actuator inputs for the proposed method at the 6m/s wind velocity of Case B from the lateral direction.

method is much superior to the traditional Dryden method under large lateral wind disturbance (>2m/s), although they could achieve similar performance under low wind speed (2m/s). Hence, the proposed position controller based on the force/moment compensation could achieve smoother and faster performance to attenuate large short-period gust wind disturbance from the longitude or lateral direction.

**Case B:** In Case B, Type II wind disturbances were applied to simulate long-lasting wind disturbances from 0 (longitude) and 270 (lateral) degree directions, respectively. Fig.12 shows the position regulation of the helicopter based on the velocity compensation where the parameters are chosen as $\alpha$ =8, $\beta$ =6 in the conditions of different longitude wind speeds (2m/s~8m/s). Fig.13 shows the position regulation result based on the proposed force/moment compensation method where the parameters are chosen as $\alpha$=3, $\beta$=2.5 in the same range of longitude wind speeds. Fig.14 shows the control inputs of the proposed force/moment compensation method when the longitude velocity of long-lasting wind disturbances is 6m/s. According to Fig. 12, the position controller based on the velocity compensation gradually departs from the desired position, especially for larger wind disturbance. When the wind velocity is up to 8m/s, the steady-state error could be beyond 3m. According to the results in Fig. 13, the position of the UAH converges to the desired position very well using the proposed force/moment compensation method.

In the conditions of different lateral wind speeds (2m/s~8m/s), Fig.15 shows the position regulation of the helicopter based on the velocity compensation where the parameters are chosen as $\alpha$=6, $\beta$=4. Fig.16 shows the position regulation result based on the proposed force/moment compensation method where the parameters are chosen as $\alpha$=2, $\beta$=2.5 under the same range of lateral wind speeds. Fig.17 shows the control inputs of the proposed force/moment compensation method when the lateral velocity of long-lasting wind disturbances is 6m/s. The position regulation performance shows pretty stable convergence using the proposed method, while the traditional velocity compensation could not even afford stronger lateral long-lasting wind. Therefore, the proposed control method with the force/moment compensation is much more robust to attenuate large horizontal wind disturbances.

Future work will involve the analysis of the system performance under the condition of different directional wind disturbances.

## 5. Conclusions

The paper presents a new hybrid control architecture to incorporate a theoretic control design and experiment-based force/moment compensation in order to attenuate large horizontal wind disturbances. The theoretic control design took advantage of the backstepping algorithm. The force/moment compensation was derived from the wind tunnel experiment data. Simulation results show that the proposed control method can well attenuate large horizontal wind disturbances to the desired position.

## 6. Acknowledgments

This research is supported by the NSF of China under NSFC Grant No. 60905052 and by the State Key Laboratory of Robotics and System (HIT) under SKLRS-2011-ZD-04.

## 7. References

[1] O. Amidi, Y. J. Mesaki, and T. Kanade, "Research on an Autonomous Vision-Guided Helicopter," *Journal of the American Institute of Aeronautics and Astronautics,* pp. 456-463, 1993.

[2] M. Sugeno, M. F. Griffin, and A. Bastian, "Fuzzy Hierarchical Control of an Unmanned Helicopter," in *Proceedings of International Fuzzy System Association Conference*, Seoul, Korea, 1993, pp. 179–182.

[3] C. Cavalcante, J. Cardoso, J. J. G. Ramos, and O. R. Neves, "Design and Tuning of a Helicopter Fuzzy Controller," in *Proceedings of the 1995 IEEE International Conference on Fuzzy Systems*, Yokohama, Japan 1995, pp. 1549-1554.

[4] G. Buskey, G. Wyeth, and J. Roberts, "Autonomous Helicopter Hover Using an Artificial Neural Network " in *2001 IEEE International Conference on Robotics and Automation*, Seoul, South Korea 2001, pp. 1635 -1640.

[5] A. Y. Ng, H. J. Kim, M. I. Jordan, and S. Sastry, "Autonomous Helicopter Flight via Reinforcement Learning," in *17th Conference on Neural Information Processing Systems*, British Columbia, Canada, 2003.

[6] P. Abbeel, A. Coates, M. Quigley, and A. Y. Ng, "An Application of Reinforcement Learning to Aerobatic Helicopter Flight," in *Proceedings of the 12th Annual Conference on Neural Information Processing Systems*, Vancouver, British Columbia, Canada, 2006, pp. 1-8.

[7] A. Coates, P. Abbeel, and A. Y. Ng, "Learning for Control from Multiple Demonstrations," in *Proceedings of the 25th International Conference on Machine Learning*, New York, NY, USA, 2008 pp. 144-151.

[8] B. Mettler, M. Tischler, and T. Kanade, "Attitude Control Optimization for a Small-Scale Unmanned Helicopter," in *AIAA Guidance, Navigation and Control Conference*, Minneapolis, USA, 2000, pp. 289-302.

[9] T. J. Koo and S. S. Sastry, "Output Tracking Control Design of a Helicopter Model Based on Approximate Linearization," in *Proceedings of the 37th IEEE Conference on Decision and Control*, Tampa, Florida, USA 1998, pp. 3635-3640

[10] B. Kadmiry, P. Bergsten, and D. Driankov, "Autonomous Helicopter Control Using Fuzzy Gain Scheduling," in *Proceedings of IEEE International Conference on Robotics and Automation*, Seoul, Korea, 2001, pp. 2980-2985.

[11] H. J. Kim and D. H. Shim, "A Flight Control System for Aerial Robots: Algorithms and Experiments," *Control Engineering Practice,* vol. 11, pp. 1389-1400, 2003.

[12] J. F. Du, T. S. Lu, Y. Zhang, G. Wang, and Z. G. Zhao, "Model Predictive Control with Application to a Small-Scale Unmanned Helicopter," in *Embedded Systems – Modeling, Technology, and Applications*, G. Hommel and S. Huanye, Eds., ed: Springer Netherlands, 2006, pp. 131-139.

[13] K. P. Tee, S. S. Ge, and F. E. H. Tay, "Adaptive Neural Network Control for Helicopters in Vertical Flight," *IEEE Transactions on Control Systems Technology*, vol. 16, pp. 753-762, 2008.

[14] B. M. Chen, S. S. Ge, and B. Ren, "Robust attitude control of helicopters with actuator dynamics using neural networks," *IET Control Theory and Applications*, vol. 4, pp. 2837-2854, 2010.

[15] M. L. Civita, G. Papageorgiou, W. Messner, and T. Kanade, "Design and Flight Testing of a High-Bandwidth H-infinity Loop Shaping Controller for a Robotic Helicopter," *Journal of Guidance, Control, and Dynamics*, vol. 29, pp. 485-494, 2006.

[16] H. Q. Wang, D. B. Wang, A. A. Mian, and H. B. Duan, "Multi-Mode Flight Control for an Unmanned Helicopter Based on Robust H∞ D-Stabilization and PI Tracking Configuration," *Space Research Journal*, vol. 1, pp. 39-52, 2008.

[17] Y. J. Xu, "Multi-Timescale Nonlinear Robust Control for a Miniature Helicopter," in *2008 American Control Conference*, Seattle, Washington, USA, 2008, pp. 2546-2551.

[18] G. W. Cai, B. M. Chen, X. X. Dong, and T. H. Lee, "Design and implementation of a robust and nonlinear flight control system for an unmanned helicopter," *Mechatronics*, vol. 21, pp. 803-820, 2011.

[19] B. Q. Song, J. K. Mills, Y. H. Liu, and C. Z. Fan, "Nonlinear Dynamic Modeling and Control of a Small-Scale Helicopter," *International Journal of Control, Automation and Systems*, vol. 8, pp. 534-543, 2010.

[20] R. Mahony, T. Hamel, and A. Dzul, "Hover Control via Lyapunov Control for an Autonomous Model Helicopter," in *Proceedings of the 38th IEEE Conference on Decision and Control*, Arizona, USA, 1999, pp. 3490-3495 vol.4.

[21] E. Frazzoli, M. A. Dahleh, and E. Feron, "Trajectory Tracking Control Design for Autonomous Helicopters Using a Backstepping Algorithm," in *Proceedings of the American Control Conference*, Chicago, Illinois 2000, pp. 4102-4107.

[22] H. R. Pota, B. Ahmed, and M. Garratt, "Velocity Control of a UAV Using Backstepping Control," in *45th IEEE Conference on Decision and Control*, San Diego, CA, USA., 2006, pp. 5894-5899.

[23] B. Ahmed, H. R. Pota, and M. Garratt, "Flight Control of a Rotary wing UAV using Adaptive Backstepping," in *2009 IEEE International Conference on Control and Automation*, Christchurch, New Zealand, 2009, pp. 1780-1785.

[24] C. T. Lee and C. C. Tsai, "Adaptive Backstepping Integral Control of a Small-Scale Helicopter for Airdrop Missions," *Asian Journal of Control*, vol. 12, pp. 531-541, 2010.

[25] J. M. Buffington, H. H. Yeh, and S. S. Banda, "Robust Control Design for an Aircraft Gust Attenuation Problem," in *Proceedings of the 31st IEEE Conference on Decision and Control*, Tucson, Arizona, USA, 1992, pp. 560-561.

[26] N. Aouf, B. Boulet, and R. Botez, "Robust Gust Load Alleviation for a Flexible Aircraft," *Canadian Aeronautics and Space Journal*, vol. 46, pp. 131-139 2000.

[27] A. Martini, F. Léonard, and G. Abba, "Dynamic Modelling and Stability Analysis of Model-Scale Helicopters Under Wind Gust," *Journal of Intelligent and Robotic Systems*, vol. 54, pp. 647-686, 2009.

[28] X. Yang, H. Pota, and M. Garratt, "Design of a Gust-Attenuation Controller for Landing Operations of Unmanned Autonomous Helicopters," in *18th IEEE International Conference on Control Applications*, Saint Petersburg, Russia, 2009, pp. 1300-1305.

[29] T. Cheviron, F. Plestan, and A. Chriette, "A robust guidance and control scheme of an autonomous scale helicopter in presence of wind gusts," *International Journal of Control*, vol. 82, pp. 2206-2220, 2009.

[30] K. A. Danapalasingam, J. J. Leth, A. La Cour-Harbo, and M. Bisgaard, "Robust helicopter stabilization in the face of wind disturbance," in *49th IEEE Conference on Decision and Control*, 2010, pp. 3832–3837.

[31] F. Leonard, A. Martini, and G. Abba, "Robust Nonlinear Controls of Model-Scale Helicopters Under Lateral and Vertical Wind Gusts," *IEEE Transactions on Control Systems Technology*, vol. 20, pp. 154-163, 2012.

[32] R. W. Prouty, *Helicopter Performance, Stability, and Control*: PWS Publishers, Boston, MA, 1986.

[33] N. Robert, *Flight Stability and Automatic Control*. New York: McGraw-Hill Companies, 1998.

[34] T. J. Koo and S. S. Sastry, "Differential Flatness Based Full Authority Helicopter Control Design," in *Proceedings of the 38th IEEE Conference on Decision and Control*, Arizona, USA, 1999, pp. 1982-1987.

[35] R. K. Heffley and M. A. Mnich, "Minimum-Complexity Helicopter Simulation Math Model," Technical Report, NASA Contractor Report 177476, Moffett Field, CA, NASA,1988.

[36] (1997). *U.S. Military Handbook MIL-HDBK-1797*. Available: http://www.everyspec.com/MIL-HDBK/MIL-HDBK+(1500+-+1799)/MIL-HDBK-1798_28506/